\documentstyle[11pt]{article}

\begin{document}

\title{\bf Orthodox Gravity} \author{\bf A.Y. Shiekh \\
International Centre for Theoretical Physics, Miramare,
Trieste, Italy}

\date{14th July 1993}

\maketitle

\begin{abstract} A scalar field theory is investigated
within the context of orthodox quantum gravity.
\end{abstract}


\section{Introduction}
  With quantum theory and general relativity being such
good descriptions of the world, it is  somewhat paradoxical
that we have still not managed to wed the two theories
[Isham, 1981].  Before embarking upon variations one might
question the need to quantize gravity at all, since  there
is no direct experimental evidence demanding the
quantization of the gravitational field  [Feynman, 1963].
However, if gravity remains classical, since its fields are
then not subject to  the uncertainty principle of quantum
theory, it might be employed to make an indirect
measurement of a quantum field that would be more precise
than that permitted. This argument  for quantizing gravity
is not watertight, as one might propose a gravitational
coupling to the  quantum expectation value, or some other
alteration to quantum theory itself [Kibble, 1981].
However it does motivate one to begin by investigating the
obstacles to naive quantization of  the gravitational field.

  The usual scheme of field quantization is plagued by
divergences, but in some special cases  those infinities
can be consistently ploughed back into the theory to yield
a finite end result with  a small number of arbitrary
constants remaining; these then being obtained from
experiment  [Ramond, 1990; Collins, 1984]. This is the
renowned scheme of renormalization, disapproved of  by
some, but reasonably well defined and yielding results in
excellent agreement with nature.  The fact that only some
theories are renormalizable has the beneficial effect of
being selective,  and so predictive. Unfortunately, in the
usual sense, general relativity is {\em not} renormalizable
[Veltman, 1976].

\subsection{Traditional Formulation}
  Orthodox quantum gravity is a perturbatively
unrenormalizable theory in the traditional sense,  for
starting from the example of a free scalar field in a
gravitational field described by the  Lagrangian:

$${\cal L}=R+{\textstyle{1 \over 2}}g^{\mu \nu}\partial
_\mu \phi \partial _\nu \phi + {\textstyle{1\over
2}}m^2\phi ^2$$

one discovers, upon quantizing both the matter and
gravitational fields, that the counter terms  do not fall
back within the original Lagrangian. Already at one loop
one observes the appearance  of $\phi^4$  and $p^4$
 counter terms (most easily seen by power counting); where
$p^2$ is shorthand for $g^{\mu \nu }\partial _\mu \phi
\partial _\nu \phi $, and not the independent variable of
Hamiltonian mechanics. At two loop one also  has such
divergences, along with the occurrence of additional
counter terms of   $\phi^6$ and $p^6$ form.  This continues
indefinitely, and since the total number of counter terms
is then infinite in  number, their associated ambiguities
destroy the predictive power of the theory. The presence
of higher derivative counter terms further destroys the
causal behaviour of the theory.

  In summary, renormalizability requires that the number of
independent counter terms be finite  in number and that
they do not spoil the physical behaviour of the original
theory (modification is  permitted). The fact that even
after successful renormalization some factors, such as mass
and  charge are left undetermined should perhaps not be
viewed as a predictive shortcoming, since  the fundamental
units of nature are relative; that is to say, the choice of
reference unit (be it  mass, length, time or charge) is
always arbitrary, and then everything else can be stated
in  terms of these few units.

  These observations motivate the consideration of the most
general Lagrangian (in even  powers of $\phi $
and $p$) permitted on the grounds of symmetry:

$$ \begin{array}{ll}
{\cal L}
&=\Lambda +R+{\textstyle{1 \over
2}}p^2+{\textstyle{1 \over 2}}m^2\phi ^2+p^2\phi ^2\kappa
(\phi ^2)+\phi ^4\lambda (\phi ^2) \\ \\
&+R\phi ^2\gamma (\phi^2)
 +\underbrace {Rp^2e(p^2,\phi ^2)+\left( {\alpha R^2+\beta
R_{\mu \nu }R^{\mu \nu }+\ldots } \right)f(p^2,\phi
^2)}_{higher\  derivative \ terms} \end{array} $$

 where $\kappa$, $\lambda$, $\gamma$, $e$, and $f$ are
arbitrary analytic functions.

  Strictly this is formal in having neglected gauge fixing
and the resulting presence of ghost  particles. Symmetry
now assures us that all counter terms must fall back within
this  Lagrangian, and it is this that motivated the
construction. However the theory in this form has  no
predictive content, since there are an infinite number of
arbitrary constants (in each arbitrary  function: $\kappa$,
$\lambda$, $\gamma$, $e$, $f$), and in this sense the theory
is not renormalized. However, there remain  physical
criterion to pin down some of these arbitrary factors.

  The cosmological constant is abandoned on the grounds of
energy conservation [Shiekh, 1992], and since in general
the higher derivative terms lead to acausal behaviour,
their renormalised  coefficient can also be put down to
zero. This still leaves the three arbitrary functions
$\kappa$, $\lambda$ and $\gamma$, associated with the terms
$p^2\phi ^2\kappa (\phi ^2)$,  $\phi ^4\lambda (\phi ^2)$
and $R\phi ^2\gamma (\phi ^2)$.
 The last may be abandoned on the grounds of defying the
equivalence principle. To see this, begin by considering
the first  term of the Taylor expansion, namely $R\phi ^2$;
this has the form of a mass term and so one would  be able
to make a local measurement of mass to determine the
curvature, and so contradict the  equivalence principle.
The same line of reasoning applies to the remaining terms,
$R\phi ^4$, $R\phi ^6$, ...  etc.

  This leaves us the two remaining infinite families of
ambiguities within the terms  $\phi ^4\lambda (\phi^2)$
and  $p^2 \phi^2 \kappa (\phi^2)$.  In the limit of flat
space in 3+1 dimensions this will reduce to a renormalized
theory  in the traditional sense if  $\lambda
(\phi^2)=constant$, and   $\kappa (\phi^2)=0$. So one is
lead to proposing that the  renormalised theory of quantum
gravity for a scalar field should have the form:

$$ \begin{array}{ll} {\cal L}_{ren}
&= \Lambda_0 + R_0 +
{\textstyle{1\over 2}} p_0^2 + {\textstyle{1 \over 2}} m_0^2
\phi_0^2 + p_0^2 \phi_0^2 \kappa_0 (\phi _0^2)+\phi_0^4
\lambda_0 (\phi_0^2) \\ \\
&+ R_0 \phi_0^2 \gamma_0 (\phi_0^2)
+R_0 p_0^2 e_0 (p_0^2,\phi _0^2) +\left( {\alpha_0 R_0^2 +
\beta_0 R_{0 \mu \nu} }  R_0^{\mu \nu }+\ldots \right)
f_0(p_0^2,\phi_0^2) \end{array} $$

where the physical parameters:

$$ \begin{array}{ll} \Lambda =\kappa (\phi ^2)= \gamma
(\phi ^2)=e(p^2,\phi^2)=f(p^2,\phi ^2)=0 \\ \lambda (\phi
^2)=\lambda = scalar\  particle \ selfcoupling \ constant \\
m = mass \ of \ the \ scalar \ particle  \end{array} $$

  One might wonder about the renormalization group
parameter. Although this would only be  one additional free
parameter, there are hints that this might be fixed
[Culumovic et al., 1990;  Leblanc et al., 1991], and one
might there anticipate the appearance of the Plank mass.

  We are left with a finite theory that has few arbitrary
constants. Despite the patch work line of  reasoning
invoked to arrive at this hypothesis, one might alter
perspective and simply be  interested in investigating the
consequences of such a scheme for its own sake, where many
of  the arbitrary factors are set to zero, for whatever
reason. At this stage any well behaved, finite  theory is
worth investigating; and it is unfortunate that we don't
have the guiding hand of  mother nature to assist us in the
guessing game.

\subsection{Viable Formulation}
  Having discussed this approach within the context of
traditional renormalization; it is intriguing  to note that
the use of analytic continuation [Bollini at al., 1964;
Speer, 1968; Salam and  Strathdee, 1975; Hawking, 1975;
Dowker and Critchley, 1976] and the more recent method of
operator regularization [McKeon and Sherry, 1987; McKeon et
al., 1987; McKeon et al., 1988;  Mann, 1988; Mann et al.,
1989; Culumovic et al., 1990; Shiekh, 1990], implements the
above  scheme in a much cleaner way.

  In operator regularization one removes divergences using
the analytical continuation:

$$H^{-m}=\mathop {\lim }\limits_{\varepsilon \to 0} {{d^n}
\over {d\varepsilon ^n}}\left( {{{\varepsilon ^n} \over {n!
}}H^{-\varepsilon -m}} \right)$$

where $n$ is chosen sufficiently large that one is without
infinities. This is explicitly illustrated  through an
example later.

  The method of operator regularization has the strength of
explicitly maintaining invariances, as  well as being
applicable to all loop levels; unlike the original Zeta
function technique [Salam and  Strathdee, 1975; Dowker and
Critchley, 1976; Hawking, 1975] that only applied to one
loop.

  To see this method in action, we will walk through a
simple example of a divergent one loop  diagram of a
massive scalar particle in quantum gravity. So begin with
an investigation of a  massive scalar theory in its own
induced gravitational field, described by the action:

$${\cal S}=\int_{-\infty }^\infty  {\sqrt g}d^4x\left(
{R+{\textstyle{1 \over 2}}g^{\mu \nu }\partial _\mu \phi
\partial _\nu \phi +{\textstyle{1 \over 2}}m^2\phi
^2+\lambda \phi ^4} \right)$$

  The Feynman rules (of which there are an infinite number)
we explicitly list; the gauged  graviton propagator being
derived from the gravitational, $R$, Lagrangian [M. Veltman,
Les  Houches XXVIII, 1976] (see FIG. 1):

$${{\delta _{\mu \alpha }\delta _{\nu \beta }+\delta _{\mu
\beta }\delta _{\nu \alpha }-\delta _{\mu \nu }\delta
_{\alpha \beta }} \over {p^2}}$$

  The scalar propagator (see FIG. 2):

$${1 \over {p^2+m^2}}$$

  First interaction vertex (see FIG. 3):

$$\kappa _{21}\left[ {{\textstyle{1 \over 2}}\delta _{\mu
\nu }\left( {p\cdot q-m^2} \right)-p_\mu q_\nu } \right]$$

etc.

  Although there are an infinite number of Feynman
diagrams, only a finite number are used to  any finite loop
order.

\subsubsection{Divergent One loop diagram example:}

  Set about a one loop investigation with matter particles
on the external legs (see FIG. 4).

$$ \begin{array}{ll} =\int_{-\infty }^\infty  {{{d^4l}
\over {\left( {2\pi } \right)^4}}}\left( {{1 \over
{l^2+m^2}}} \right)\left( {{{\delta _{\mu \alpha }\delta
_{\nu \beta }+\delta _{\mu \beta }\delta _{\nu \alpha
}-\delta _{\mu \nu }\delta _{\alpha \beta }} \over {\left(
{l+p} \right)^2}}} \right)
  \kappa _{21}\left[ {{\textstyle{1 \over 2}}\delta _{\mu
\nu }\left( {p\cdot l-m^2} \right)-p_\mu l_\nu }
\right]\kappa _{21}\left[ {{\textstyle{1 \over 2}}\delta
_{\alpha \beta }\left( {p\cdot l-m^2} \right)-p_\alpha
l_\beta } \right] \end{array} $$

expand out the indices to yield:

$$=\kappa _{21}^2\int_{-\infty }^\infty  {{{d^4l} \over
{\left( {2\pi } \right)^4}}}{1 \over {l^2+m^2}}{1 \over
{\left( {l+p} \right)^2}}\left( {p^2l^2+2m^2p\cdot l-2m^4}
\right)$$

  Then introduce the standard Feynman parameter 'trick':

$${1 \over {D_1^{a_1}D_2^{a_2}\ldots D_k^{a_k}}}={{\Gamma
(a_1+a_2+\ldots a_k)} \over {\Gamma (a_1)\Gamma (a_2)\ldots
\Gamma (a_k)}}\int_0^1 \ldots \int_0^1 {dx_1\ldots
dx_k}{{\delta (1-x_1-\ldots x_k)x_1^{a_1-1}\ldots
x_k^{a_k-1}} \over {\left( {D_1x_1+\ldots D_kx_k}
\right)^{a_1+\ldots a_k}}}$$

to yield:

$$=\kappa _{21}^2\int_{-\infty }^\infty  {{{d^4l} \over
{\left( {2\pi } \right)^4}}\int_0^1
{dx}}{{p^2l^2+2m^2p\cdot l-2m^4} \over {\left[
{l^2+m^2x+p^2\left( {1-x} \right)+2l\cdot p\left( {1-x}
\right)} \right]^2}}$$

  Remove divergences using the analytic continuation:

$$H^{-m}=\mathop {\lim }\limits_{\varepsilon \to 0}{{d^n}
\over {d\varepsilon ^n}}\left( {{{\varepsilon ^n} \over
{n! }}H^{-\varepsilon -m}} \right)$$

  $n$ being chosen sufficiently large to cancel the
infinities. For the case in hand $n=1$ is adequate.

$$H^{-2}=\mathop {\lim }\limits_{\varepsilon \to 0}{d \over
{d\varepsilon }}\left( {\varepsilon H^{-\varepsilon -2}}
\right)$$

  This yields:

$$=\kappa _{21}^2\int_0^1 {dx}\;\mathop {\lim
}\limits_{\varepsilon \to 0}{d \over {d\varepsilon
}}\int_{-\infty }^\infty  {{{d^4l} \over {\left( {2\pi }
\right)^4}}}\left( {\varepsilon {{p^2l^2+2m^2p\cdot l-2m^4}
\over {\left[ {l^2+m^2x+p^2\left( {1-x} \right)+2l\cdot
p\left( {1-x} \right)} \right]^{\varepsilon +2}}}}
\right)$$

  Then performing the momentum integrations using [Ramond,
1990]:

$$\int_{-\infty }^\infty  {{{d^{2\omega }l} \over {\left(
{2\pi } \right)^{2\omega }}}}{1 \over {\left(
{l^2+M^2+2l\cdot p} \right)^A}}={1 \over {(4\pi )^\omega
\Gamma (A)}}{{\Gamma (A-\omega )} \over
{(M^2-p^2)^{A-\omega }}}$$

$$\int_{-\infty }^\infty  {{{d^{2\omega }l} \over {\left(
{2\pi } \right)^{2\omega }}}}{{l_\mu } \over {\left(
{l^2+M^2+2l\cdot p} \right)^A}}=-{1 \over {(4\pi )^\omega
\Gamma (A)}}p_\mu {{\Gamma (A-\omega )} \over
{(M^2-p^2)^{A-\omega }}}$$

$$\int_{-\infty }^\infty  {{{d^{2\omega }l} \over {\left(
{2\pi } \right)^{2\omega }}}}{{l_\mu l_\nu } \over {\left(
{l^2+M^2+2l\cdot p} \right)^A}}={1 \over {(4\pi )^\omega
\Gamma (A)}}\left[ {p_\mu p_\nu {{\Gamma (A-\omega )} \over
{(M^2-p^2)^{A-\omega }}}+{{\delta _{\mu \nu }} \over
2}{{\Gamma (A-\omega -1)} \over {(M^2-p^2)^{A-\omega -1}}}}
\right]$$

yields the finite expression:

$$={{\kappa _{21}^2} \over {\left( {4\pi }
\right)^2}}\int_0^1 {dx}\;\mathop {\lim
}\limits_{\varepsilon \to 0}{d \over {d\varepsilon }}\left(
\matrix{{{p^4(1-x)^2\Gamma (\varepsilon )} \over {\left[
{m^2x+p^2x\left( {1-x} \right)} \right]^\varepsilon
}}+2{{p^2\Gamma (\varepsilon -1)} \over {\left[
{m^2x+p^2x\left( {1-x} \right)} \right]^{\varepsilon
-1}}}\cr
  -2{{m^2p^2(1-x)\Gamma (\varepsilon )} \over {\left[
{m^2x+p^2x\left( {1-x} \right)} \right]^\varepsilon
}}-2{{m^4\Gamma (\varepsilon )} \over {\left[
{m^2x+p^2x\left( {1-x} \right)} \right]^\varepsilon }}\cr}
\right){\varepsilon  \over {\Gamma (\varepsilon
+2)}}$$

Doing the $\varepsilon$ differential using:

$$\mathop {\lim }\limits_{\varepsilon \to 0}{d \over
{d\varepsilon }}\left( {{\varepsilon  \over {\Gamma
(\varepsilon +2)}}\left( {a{{\Gamma (\varepsilon )} \over
{\chi ^\varepsilon }}+b{{\Gamma (\varepsilon -1)} \over
{\chi ^{\varepsilon -1}}}} \right)} \right)=-a\ln (\chi
)+b\chi \left( {\ln (\chi )-1}
\right)$$

yields:

$$={{\kappa _{21}^2} \over {\left( {4\pi }
\right)^2}}\int_0^1 {dx}\left( \matrix{\left( {\left(
{2m^4+2m^2p^2-p^4} \right)+p^4x\left( {4-3x} \right)}
\right)\ln \left( {m^2x+p^2x(1-x)} \right)\cr
  -2p^2(m^2x+p^2x(1-x))\cr} \right)$$

and finally performing the $x$ integration gives rise to the
final result:

$$={{\kappa _{21}^2} \over {\left( {4\pi }
\right)^2}}m^4\left( {\left( {3+2{\textstyle{{p^2} \over
{m^2}}}+{\textstyle{{m^2} \over {p^2}}}} \right)\ln
(1+{\textstyle{{p^2} \over {m^2}}})-3-{\textstyle{9 \over
2}}{\textstyle{{p^2} \over {m^2}}}-{\textstyle{1 \over
6}}{\textstyle{{p^4} \over {m^4}}}+2\left(
{1+{\textstyle{{p^2} \over {m^2}}}}
\right)ln({\textstyle{{m^2} \over {\mu ^2}}})}
\right)$$

where there is no actual divergence at $p=0$, and it should
be commented that the use of a  computer mathematics
package can in general greatly reduced 'calculator'
fatigue. The factor $\mu$ appears on dimensional grounds.

\section{Acknowledgements}

  I should like to thank John Strathdee and Seifallah
Randjbar Daemi for listening to and  commenting upon, if
not necessarily agreeing with, my ideas. For lending a
responsive ear;  K. Stelle, G. 't Hooft and D. McKeon.

\section{References}

  The field is so well explored that rather than attempt
any kind of complete referencing, a  flavour is given.
\\
\\
{\noindent
\footnotesize
1)	C.J. Isham,
\\{\em 'Quantum Gravity - An
Overview'}, in "Quantum Gravity 2: A Second Oxford
Symposium" pp. 1-62,  eds. C.J. Isham, R. Penrose and D.W.
Sciama, (Oxford University Press, Oxford, 1981).
\\
\\
2)	R.P.
Feynman,
\\{\em "Lectures on Gravitation"} (Caltech,
1962-1963, unpublished).
\\
\\
3)	T.W.B. Kibble,
\\{\em 'Is a
Semi-Classical Theory of Gravity Viable?'}, in "Quantum
Gravity 2: A Second Oxford  Symposium" pp. 63-80, eds. C.J.
Isham, R. Penrose and D.W. Sciama, (Oxford University
Press, Oxford, 1981).
\\
\\
4)	P. Ramond,
\\{\em "Field Theory:
A Modern Primer"}, 2nd Ed (Addison-Wesley, 1990).
\\
5)	J. Collins,
\\{\em "Renormalization"}, Cambridge
University Press, London, 1984.
\\
\\
6)	M. Veltman,
\\{\em 'Quantum Theory of Gravitation'}, Les Houches XXVIII,
"Methods In Field Theory", pp. 265- 327, eds. R. Ballan and
J. Zinn-Justin, (North-Holland, Amsterdam, 1976).
\\
\\
7)	A. Shiekh,
\\{\em 'Does Nature place a Fundamental Limit on
Strength?'}, Can.J.Phys., {\bf 70}, 1992, 458-
\\
\\
8)	M. Leblanc, R.B. Mann, D.G.C. McKeon and T.N. Sherry,
\\{\em 'The Finite Effective Action in the Non-Linear Sigma  Model
with Torsion to Two-Loop Order'}, Nucl.Phys. {\bf B349},
1991, 494-
\\
9)	L. Culumovic, D.G.C. McKeon and T.N. Sherry,
\\{\em 'Operator regularization and massive Yang-Mills theory'},
Can.J.Phys., {\bf 68}, 1990, 1149-
\\
\\
10)	C. Bollini, J.
Giambiagi and A. Dominguez,
\\{\em 'Analytic Regularization
and the Divergences of Quantum Field  Theories'}, Nuovo
Cimento {\bf 31}, 1964, 550-
\\
11)	E. Speer,
\\{\em 'Analytic Renormalization'},
J.Math.Phys. {\bf 9}, 1968, 1404-
\\
\\
12)	A. Salam and J.
Strathdee, Nucl.Phys. {\bf B90},
\\{\em 'Transition
Electromagnetic Fields in Particle Physics'}, 1975, 203-
\\
13) J. Dowker and R. Critchley,
\\{\em 'Effective Lagrangian
and energy-momentum tensor in de sitter space'},
Phys.Rev., {\bf D13}, 1976,
3224-  \\
14)	S. Hawking,
\\{\em 'Zeta Function Regularization of Path
Integrals in Curved Space'}, Commun.Math.Phys., {\bf 55},
1977,  133-
\\
\\
15)	D. McKeon and T. Sherry,
\\{\em 'Operator
Regularization of Green's Functions'}, Phys.Rev.Lett.,
{\bf 59}, 1987, 532-
\\
16) D. McKeon and T. Sherry,
\\{\em 'Operator Regularization
and one-loop Green's functions'}, Phys.Rev., {\bf D35},
1987, 3854-
\\
17)	D. McKeon, S. Rajpoot and T. Sherry,
\\{\em 'Operator
Regularization with Superfields'}, Phys.Rev., {\bf D35},
1987, 3873-
\\
18)	D. McKeon, S. Samant and T. Sherry,
\\{\em 'Operator
regularization beyond lowest order'}, Can.J.Phys., {\bf
66}, 1988, 268-
\\
19)	R. Mann.,
\\{\em 'Zeta function regularization of Quantum
Gravity'}, In Proceedings of the CAP-NSERC Summer  Workshop
on Field Theory and Critical Phenomena. Edited by G.
Kunstatter, H. Lee, F. Khanna and H. Limezawa,  World
Scientific Pub. Co. Ltd., Singapore, 1988, p. 17.
\\
20)	R. Mann, D. McKeon, T. Steele and T. Tarasov,
\\{\em
'Operator Regularization and Quantum Gravity'}, Nucl.Phys.,
{\bf B311}, 1989, 630-
\\
21)	L. Culumovic, M. Leblanc, R. Mann, D. McKeon and T.
Sherry,
\\{\em 'Operator regularization and multiloop
Green's  functions'}, Phys. Rev., D41, 1990, 514- 22)	A.
Shiekh, Can.J.Phys., 'Zeta-function regularization of
quantum field theory', {\bf 68}, 1990, 620- }

\end{document}